\documentclass[10pt]{iopart}
\begin{document}
\title[Second variation of the Helfrich-Canham Hamiltonian...]{Second
variation of the
Helfrich-Canham Hamiltonian  and reparametrization invariance} 
\author{R  Capovilla\dag and
J Guven\ddag\S}
\address{\dag\
Departamento de F\'{\i}sica,
 Centro de Investigaci\'on
y de Estudios Avanzados del IPN,
Apdo Postal 14-740, 07000 M\'exico,
D. F.,
MEXICO}
\address{\ddag\ School of Theoretical Physics,
Dublin Institute for Advanced Studies,
10 Burlington Road, Dublin 4, IRELAND} 
\address{\S\
Instituto de Ciencias Nucleares,
 Universidad Nacional Aut\'onoma de M\'exico,
 Apdo. Postal 70-543, 04510 M\'exico, DF, MEXICO}

\begin{abstract}
A covariant approach towards a theory of deformations is developed to examine
both the first and second variation of  the Helfrich-Canham
Hamiltonian --- quadratic in extrinsic curvature
--- which describes fluid vesicles at mesoscopic scales.
Deformations are decomposed into tangential and normal components;
At first order, tangential deformations may always be identified with a
reparametrization; at second order, they differ. The
relationship between tangential deformations and reparametrizations,
as well as the coupling between tangential and normal deformations,
is examined at this order for both the metric 
and the extrinsic curvature tensors.
Expressions for the expansion to second order in deformations of
geometrical invariants constructed with these tensors are obtained;
in particular, the expansion of the Hamiltonian to this order
about an equilibrium is considered. Our approach applies as well
to any geometrical model for membranes.
\end{abstract}


\section{ Introduction}

In water, lipid molecules assemble spontaneously into vesicles
which are described remarkably well at mesoscopic scales by a
purely geometrical Hamiltonian \cite{Saf:94,Handbook,Peliti}. On such
scales, there is a
difference of several orders of magnitude between the thickness
of the lipid bilayer and the diameter of the vesicle; it is
therefore sensible to describe the vesicle itself as a
two-dimensional surface, the relevant course grained degrees of
freedom are purely geometrical and describe the shape of this
surface. Furthermore, this membrane acts like a two dimensional
fluid: there is no cost in energy associated with tangential
displacements of the lipid constituents which preserve the area,
and thus its shear modulus vanishes. In this respect, the
membrane differs completely in its behaviour
from a familiar elastic solid. As a two-dimensional fluid the membrane is then
described by an effective energy that does not penalize
tangential displacements. Infinitesimally, tangential
displacements can be identified with a reparametrization of the
surface. The appropriate Hamiltonian must therefore be a
geometrical  invariant under reparametrizations.

The Helfrich-Canham Hamiltonian quadratic in the mean extrinsic
curvature describes the penalty associated with the bending of
the vesicle \cite{Can,Hel,Evans}. The microscopic physics of the
lipid molecules is encoded in the rigidity modulus characterizing
the stiffness of the membrane on the particular mesoscopic scale being considered.

There are global constraints on the shape of the vesicle: the area is
fixed, and on time scales relevant in experiments, the enclosed
volume is also. At first order, the particular composition of the
lipid bilayer, and in particular the asymmetry between the
layers, is characterized  by a constraint or a penalty on the total
mean extrinsic curvature of the surface, a quantity which captures the area
difference between the two layers. For the sake of simplicity we
will restrict our attention to a minimal geometric model for
fluid vesicles which is known as the strict bilayer couple model
\cite{Sve.Zek:89}.
A more realistic geometric model, the area-difference model,
takes the bilayer composition into account more precisely, in
particular, the difference in stretching of the individual layers
\cite{Sve85,Boz92,Mi94}. The two models are related by a Legendre
transformation; the formal questions we  address  apply to both.

Our aim in this paper is to provide an approach  towards a {\it covariant} 
theory of deformations of a membrane described by the Helfrich-Canham
Hamiltonian, although our considerations will not depend on the
details of this model; they apply equally well to any
reparametrization invariant geometrical theory of membranes. The
basic variables are the shape functions describing the
surface. We examine how the geometry of this surface changes under
a deformation. The approach we will adopt complements the one presented
in  \cite{Defo} where the deformation was decomposed into its
tangential and normal components with a focus on the latter;
it is simliar, at least in spirit, to the approach taken by Cai
and Lubensky in their description of membrane dinamycs \cite{CL1,CL2}.

At first order in the deformation, the change in the Hamiltonian
vanishes when the vesicle is in equilibrium. At this order, the
tangential deformation of the Hamiltonian appears only in a
boundary term, so that for the purposes of examining the
equilibria of closed vesicles, it can always be neglected. This
is consistent with our understanding that an infinitesimal
tangential deformation is a reparametrization of the surface.
Contrary to what one might expect, however, this identification
breaks down at higher orders. Finite tangential deformations are
{\it not} simple exponentials of infinitesimal reparametrizations.
Nonetheless, as was shown in \cite{Defo}, if one is interested
only in fluctuations about equilibrium (so that the
Euler-Lagrange equation is satisfied), tangential deformations
remain irrelevant at second order. This is because,
at this order, the tangential contribution to the deformation
of any given term appearing in the Hamiltonian is
proportional to the Euler-Lagrange derivative of that term.
In equilibrium the sum of these terms vanishes:
the second variation of the total Hamiltonian about equilibrium is
thus always a quadratic in the normal deformation.
This is the principal justification for the
cavalier approach adopted in \cite{Defo} where tangential
deformations are discarded from the outset; considering the
effort one must expend to keep track of tangential contributions,
the fact that no error is incurred  represents a stroke of good luck.
What it fails to do, however, even at second order
is provide a correct expansion of
individual geometrical tensors, such as the
metric and the extrinsic curvature: tangential deformations not
only contribute but also couple non-trivially to normal deformations and there is no
justification to drop them.

In this paper, we will examine the coupling between tangential and
normal components at second and higher orders. We will also attempt to
quantify the extent to which
tangential deformations differ from reparametrizations.
These issues do not appear to have been addressed before.
Besides their value in point of principle, there is also a
practical value to understanding them: there are occasions
when it is necessary to look beyond second order.
To identify the stable deformations of a spherical equilibrium shape one
needs to expand the
Hamiltonian out to fourth order \cite{Hel.OuY:87,Hel.OuY:89} in order to
resolve a degeneracy occurring at second order. Helfrich and Ou-Yang  did
not attempt the full calculation focusing instead
on a single mode. Whereas at second order they can be ignored,
tangential deformations will contribute at higher orders to perturbations
about an equilibrium, and thus they must be confronted.
It is perhaps not too surprising that, to
date, a renormalization group analysis of the fluid membrane model
has not been attempted at two-loops. The number of terms
involved, even in a straighforward Monge representation of
deformations, is sufficient to discourage the feint hearted.
To attempt such an exercise, with some hope of successfully completing it,
it is important to identify underlying patterns in the expansion
of the Hamiltonian.

This article is organized as follows. Sect. 2 introduces the
geometry of two-dimensional surfaces and the Helfrich-Canham
Hamiltonian for lipid membranes. Sects. 3 and 4 consider how the
intrinsic and extrinsic geometries of the surface change under an
infinitesimal covariant deformation up to second order,
respectively. In Sect. 5, the deformation is decomposed in its
normal and tangential components and we examine the relationship
between tangential deformations and reparametrizations both at
first and second order. In Sect. 6 we derive the first order
variation of the Helfrich-Canham Hamiltonian and we identify its
Euler-Lagrange derivatives, obtaining the shape equation which
determines the equilibrium configurations. In Sect. 7, we derive
general expressions for the stresses and torques associated with
the Hamiltonian, providing an alternative derivation of the
results of  \cite{Stress}. The second order variation of the
Hamiltonian is derived in Sect. 8. We conclude in Sect. 9 with
some final remarks.

\section{ Geometric model}

We model a lipid vesicle as a 2-dimensional surface $\Sigma$ embedded in 3-dimensional
space. This surface is specified locally
in parametric form by three shape functions
${\bf X} = ( X^1,X^2,X^3)$,
$ {\bf x} = {\bf X} (\xi^a )$,
where the coordinates ${\bf x} = x^\mu = (x^1 , x^2 , x^3 )$
describe a point in space,
$\xi^a = (\xi^1 , \xi^2 )$ are arbitrary coordinates on the surface.

First, we recall briefly some basic facts about the geometry of surfaces.
For a thorough introduction to this subject see {\it e.g.}
\cite{Spivak}, \cite{doCarmo}.
The two tangent vectors to $\Sigma$ are
${\bf e}_a = \partial_a {\bf X}$, with $\partial_a =
\partial / \partial \xi^a$. The metric induced on $\Sigma$
by the embedding  is defined by
$ g_{ab} = {\bf e}_a \cdot {\bf e}_b $.
Latin indices are lowered and raised with $g_{ab}$ and its inverse
$g^{ab}$, respectively. The induced metric defines the infinitesimal
area element with $dA = \sqrt{g} \; d^2 \xi $, where $g$ denotes the
determinant of $g_{ab}$.
The unit normal to the surface $\Sigma$, ${\bf n}$, is
defined implicitly by
$ {\bf n} \cdot {\bf e}_a = 0$, and ${\bf n}\cdot {\bf n} = 1$.
We note that the basis vectors $\{ {\bf e}_a , {\bf n} \}$
are complete: given any two vectors ${\bf U}$ and ${\bf V}$,
\begin{equation}
{\bf U} \cdot {\bf V} =
({\bf U} \cdot {\bf n}) ( {\bf V} \cdot {\bf n}) +
({\bf U} \cdot {\bf e}_a) ( {\bf V} \cdot {\bf e}^a )\,.
\label{eq:comple}
\end{equation}
The classical Gauss-Weingarten equations describe the
the expansion of the surface gradients of the basis $\{ {\bf e}_a , {\bf n} \}$
adapted to the surface $\Sigma$ in terms of the basis:
\begin{eqnarray}
\partial_a \; {\bf e}_b &=& \Gamma_{ab}^c \; {\bf e}_c - K_{ab} \; {\bf n}\,,
\label{eq:gw1}
\\
\partial_a {\bf n} &=& K_{ab} \; g^{bc} \; {\bf e}_c\,.
\label{eq:gw2}
\end{eqnarray}
Here $\Gamma^c_{ab}$ denotes the Christoffel symbols of the
$\Sigma$ covariant derivative compatible with $g_{ab}$, such that
for an arbitrary surface vector $V^b$ we have $\nabla_a V^b =
\partial_a V^b + \Gamma^b_{ac} V^c $. By compatible we mean
$\nabla_a g_{bc} = 0$. The Christoffel symbols $\Gamma^c_{ab}$
are given in terms of the induced metric by
\begin{equation}
\Gamma^a{}_{bc} = {\bf e}^a \cdot \partial_b {\bf e}_c
= {1 \over 2} g^{ad } (\partial_b g_{cd} +
\partial_c g_{bd} - \partial_d g_{bc} )\,.
\end{equation}
Geometrically, the $\Gamma^c_{ab}$ are  purely intrinsic:
they depend only on the induced metric $g_{ab}$.
In a geometrical covariant description of the
surface, the Christoffel symbols  appear only through the covariant
derivative. The intrinsic Riemann curvature tensor of $\Sigma$
quantifies the degree of failure of
the covariant derivative $\nabla_a$ to commute,
\begin{equation}
(\nabla_a \nabla_b - \nabla_b \nabla_a ) V^c =
{\cal R}^c{}_{dab} V^d\,.
\label{eq:riemannder}
\end{equation}
In terms of the Christoffel symbols, the Riemann tensor
takes the form
\begin{equation}
{\cal R}^a{}_{bcd} = \partial_c \Gamma^a_{db} -
\partial_d \Gamma^a_{cb} + \Gamma^a_{ce} \Gamma^e{}_{db} - \Gamma^a_{de}
 \Gamma^e_{cb}\,.
\label{eq:riemann}
\end{equation}
Contraction of the Riemann tensor gives the
Ricci tensor ${\cal R}_{ab} = {\cal R}^c{}_{acb}$,
and the scalar curvature  is given by contraction with the
contravariant metric
${\cal R} = g^{ab} {\cal R}_{ab}$.
For a two-dimensional surface, the Riemann tensor is
completely determined by the scalar curvature
\begin{equation}
{\cal R}_{abcd} = ({\cal R} / 2 ) \; ( g_{ac} g_{bd} - g_{ad} g_{bc})\,,
\end{equation}
which implies, in particular,
${\cal R}_{ab} = (1 / 2 ) {\cal R} g_{ab}$. The scalar curvature
of a two-dimensional surface is twice the Gaussian curvature
${\rm G}$, {\it i.e.}
${\cal R} = 2 {\rm G}$.

The extrinsic curvature tensor of the surface $\Sigma$ is
\begin{equation}
K_{ab} = - {\bf n} \cdot \partial_a {\bf e}_b = K_{ba}\,.
\end{equation}
As a real symmetric two by two matrix it can always be
diagonalized. In particular, the eigenvalues $c_1, c_2$  of the
matrix $K_a{}^b = g^{bc} K_{ca}$ are the principal curvatures of
the surface. The trace  with the contravariant metric $ K =
g^{ab} K_{ab} $ is the mean curvature of the surface. With respect
to the principal curvatures $K = c_1 + c_2 $. In the literature,
often the mean curvature is denoted by $H= (1/2) K $. The
Gaussian curvature is given in terms of the principal curvatures
by their product, ${\rm G} = c_1 \; c_2 $.

The intrinsic and extrinsic geometries of $\Sigma$
are related by the Gauss-Codazzi-Mainardi equations,
which arise as integrability conditions for the Gauss-Weingarten
equations (\ref{eq:gw1}), (\ref{eq:gw2}),
\begin{eqnarray}
{\cal R}_{abcd} - K_{ac} K_{bd} + K_{ad} K_{bc} &=& 0\,,
\label{eq:g1}
\\
\nabla_a K_{bc} - \nabla_b K_{ac} &=& 0\,.
\label{eq:cm1}
\end{eqnarray}
We will use extensively their contractions with  the contravariant metric
$g^{ab}$:
\begin{eqnarray}
{\cal R}_{ab} - K K_{ab} + K_{ac} K_{b}{}^c &=& 0\,,
\label{eq:g2}
\\
{\cal R} - K^2 + K_{ab} K^{ab} &=& 0\,,
\label{eq:g3}
\\
\nabla_b K_{a}{}^b - \nabla_a K &=& 0\,.
\label{eq:cm2}
\end{eqnarray}
Note that for a 2-dimensional surface the contracted
Gauss-Codazzi equation (\ref{eq:g3}) contains the same
information as (\ref{eq:g1}).

The  fluid state of the lipid vesicle implies that in an
effective mesoscopic description shear is negligible, therefore
the vesicle Hamiltonian has to be
invariant under reparametrizations. Moreover, as recognized long
ago the  important mode of deformation is out of the surface,
corresponding to a bending \cite{Can,Hel,Evans}. The bending energy is
quadratic in the mean extrinsic curvature
\begin{equation}
F_b = \alpha \int dA \; K^2\,,
\label{eq:bending}
\end{equation}
where the constant $\alpha$ denotes the bending rigidity.
At the same order, we have also the Gaussian bending
energy
\begin{equation}
F_G = \alpha_G \int dA \; {\cal R}\,,
\label{eq:gaussian}
\end{equation}
with $\alpha_G$ the Gaussian bending rigidity.
However, if the surface has no boundary,
by the Gauss-Bonnet
theorem, the Gaussian bending energy
is a topological invariant (see {\it e.g.} \cite{doCarmo}):
\begin{equation}
F_G = 8 \pi \alpha_G (1 - {\rm g})\,,
\label{eq:gb}
\end{equation}
where ${\rm g}$ is the genus of the surface.
As such it does not contribute to the determination
of the equilibrium configurations of the membrane.
Note that at the same order we have also the geometrical invariant
$\int dA \; K^{ab} K_{ab}$. However, it is not independent
since it is related to the bending and Gaussian bending energies
via the Gauss-Codazzi equation (\ref{eq:g3}).

The lipid vesicle is subject to various geometric constraints.
The low solubility of the lipid molecules implies that its area
$A$ is constant. The low permeability of the membrane implies that
the enclosed volume $V$ is constant. We write the enclosed volume
$V$ as a surface integral with
\begin{equation}
V = {1 \over 3} \int dA \; {\bf n} \cdot {\bf X}\,.
\label{eq:vol}
\end{equation}
The bilayer architecture of the lipid membrane is captured, in a  first
approximation, by a constraint on the area difference between the layers.
This is expressed as constant total mean curvature (see \cite{Sve.Zek:89})
\begin{equation}
M = \int dA \; K\,,
\label{eq:total}
\end{equation}
since, as we shall see below,  the difference in area is proportional to the
mean extrinsic curvature.

Therefore we are led to consider the Helfrich-Canham Hamiltonian
\begin{equation}
F = F_b + \mu \; A + \beta \; M - P \; V\,,
\label{eq:model}
\end{equation}
where $\mu , P , \beta $ are the Lagrange multipliers
that enforce the constraints of constant area, volume, and total mean
extrinsic curvature, or constant area difference, respectively.

It is important to emphasize that a more realistic model of the lipid vesicle
which takes the bilayer architecture more accurately into account
is the area-difference model \cite{Sve85,Boz92,Mi94}. A thorough
discussion of the extant curvature models for lipid membranes can
be found {\it e.g.} in the reviews
\cite{Peliti,Sve.Zek:96,Seifert}.

\section{ Deformations of the intrinsic geometry}

Let us consider now deformations of the surface $\Sigma$.
A one-parameter infinitesimal deformation of the shape
functions ${\bf X} (\xi^a )$ can be described by
\begin{equation}
\widetilde{\bf X} (\xi^a )
= {\bf X} ( \xi^a ) + \epsilon {\bf W} (\xi^a
)\,;
\label{eq:defo}
\end{equation}
${\bf W} (\xi^a )$ is an arbitrary vector field, and the constant
$\epsilon$ an infinitesimal parameter. Such a deformation gives
us  a new surface $\widetilde\Sigma$. We begin by examining how
the intrinsic geometry of the two surfaces $\widetilde{\Sigma}$
and  $\Sigma$ are connected up to second order in $\epsilon$. A
tilde will be used to denote the geometrical quantities that
characterize $\widetilde\Sigma$. The content of this section can
be found in many monographs on differential geometry. However, it
is often presented in an abstract notation quite unfamiliar to
the working physicist. For this reason, we offer a self-contained
derivation of the relationship between the geometries of $\Sigma$
and $\widetilde\Sigma$.

We will not consider deformations in which ${\bf W}$ itself
depends explicitly on ${\bf X}$, such as rigid rotations of the
surface or contexts where it is useful to tie the surface local
coordinates $\xi^a$ to the embedding itself. An example of the
latter is to parametrize the surface by arclength along some
priviledged direction. Consistency would require then that the
coordinates themselves suffer a deformation. Failure to account
for this fact is a source of frequent errors in the literature.

At {\it fixed} values of the arbitrary surface coordinates $\xi^a$,
the
tangent vectors to $\Sigma$ and $\widetilde\Sigma$ are related by
\begin{equation}
\widetilde{\bf e}_a = {\bf e}_a + {\bf e}_{a\, (1)} =
 {\bf e}_a + \epsilon {\bf W}_a\,,
\label{eq:tan}
\end{equation}
where we define ${\bf W}_a = \partial_a {\bf W}$. The change in
the tangent vectors is only linear in $\epsilon$, like the shape
functions ${\bf X}$ themselves. For a constant deformation, ${\bf
W} = {\bf a} = $ const., the tangent vectors coincide. A
translation of the surface will not change its geometry.

It follows, using
 (\ref{eq:tan}), that the induced metric on $\widetilde\Sigma$
takes the form
\begin{equation}
\widetilde{g}_{ab} = \widetilde{\bf e}_a \cdot \widetilde{\bf e}_b =
g_{ab} + 2 \epsilon ( {\bf e}_{(a} \cdot  {\bf W}_{b)} )
+ \epsilon^2 ( {\bf W}_a \cdot {\bf W}_b )\,,
\label{eq:gab}
\end{equation}
where the round brackets enclosing indices denote symmetrization,
{\it i.e.} $A_{(ab)} = (1/2) (A_{ab} + A_{ba} )$. It should be
emphasized that this expression is valid to all orders in
$\epsilon$; it terminates at second order.

The area measure on $\widetilde\Sigma$ is $\widetilde{dA} =
\sqrt{\widetilde{g}} d^2 \xi$, with $\widetilde{g}$ the
determinant of $\widetilde{g}_{ab}$. It is related to the area
measure on $\Sigma$ by the expression
\begin{eqnarray}
\fl
\sqrt {\widetilde{g}} = \sqrt{g}
\{ 1 &+& \epsilon ( {\bf e}_a \cdot {\bf W}^a )
+ {\epsilon^2 \over 2} [ ( {\bf n} \cdot {\bf W}^a )( {\bf n} \cdot {\bf
W}_a )
\nonumber \\ \fl
&+& ( {\bf e}_a \cdot {\bf W}^a )( {\bf e}_b \cdot {\bf W}^b )
- ( {\bf e}_b \cdot {\bf W}^a )
( {\bf e}_a \cdot {\bf W}^b ) ] \} + {\cal O} (\epsilon^3 )\,.
\label{eq:sqrt}
\end{eqnarray}
Note that the two terms on the second line have the structure of a
determinant for the 2-dimensional matrix $({\bf e}_a \cdot {\bf
W}^b)$. In order to derive this expression,  we  need the inverse
induced metric $\widetilde{g}^{ab}$, defined by
$\widetilde{g}_{ac} \widetilde{g}^{bc} = \delta^b_a$, together
with  (\ref{eq:gab}). For this purpose, we expand
$\widetilde{g}^{ab}$ and $\widetilde{g}_{ab}$ in powers of
$\epsilon$. Collecting terms linear in $\epsilon$, we have the
condition $g^{ab}{}_{(1)} g_{bc} + g^{ab} \; g_{bc\, (1)} = 0$,
where the number in parenthesis refers to the order in $\epsilon$.
This gives
\begin{equation}
g^{ab}{}_{(1)}
= -  2 \epsilon ({\bf e}^{(a} \cdot {\bf W}^{b)})\,.
\end{equation}
At second order in $\epsilon$, we have the condition
$g^{ab}{}_{(2)} g_{bc} +
g^{ab}{}_{\, (1)} g_{bc\, (1)}
+ g^{ab} g_{bc\, (2)} = 0 $,
which, in turn,  yields
\begin{equation}
\fl
g^{ab}{}_{(2)} = \epsilon^2 [
({\bf e}^a \cdot {\bf W}_c) ({\bf e}^b \cdot {\bf W}^c ) +
 2 ( {\bf e}^{(a} \cdot {\bf W}_c )( {\bf W}^{b)} \cdot {\bf e}^c )
- ( {\bf n} \cdot {\bf W}^a )( {\bf n} \cdot {\bf W}^b )] \,,
\label{eq:gabi2}
\end{equation}
where  we have used the
completeness relation (\ref{eq:comple}) to get
$ ({\bf e}_c \cdot {\bf W}^a )
({\bf e}^c \cdot {\bf W}^b ) =
({\bf W}^a \cdot {\bf W}^b  )
- ({\bf n} \cdot {\bf W}^a ) ({\bf n} \cdot {\bf W}^b )$.
It follows that for the inverse induced metric on $\widetilde\Sigma$ we
have
\begin{eqnarray}
\fl
\widetilde{g}^{ab} =
g^{ab} &-& 2 \epsilon ( {\bf e}^{(a} \cdot  {\bf W}^{b)} )
+ \epsilon^2 [( {\bf e}^a \cdot {\bf W}_c )( {\bf e}^b \cdot {\bf W}^c )
-  ( {\bf n} \cdot {\bf W}^a )( {\bf n} \cdot {\bf W}^b )
\nonumber \\ \fl
&+& 2
( {\bf e}^{(a} \cdot {\bf W}_c )( {\bf W}^{b)} \cdot {\bf e}^c )]
+ {\cal O} (\epsilon^3 ) \,.
\label{eq:gabi}
\end{eqnarray}
Note that $\widetilde{g}^{ab}$, unlike $\tilde g_{ab}$, has corrections at
all
orders in $\epsilon$.

We now compute $\sqrt{\widetilde{g}}$ via a Taylor expansion
in $\epsilon$,
\begin{eqnarray}
\sqrt{\widetilde{g}} &=&
\sqrt{g} + \sqrt{g}_{(1)} + \sqrt{g}_{(2)}  + {\cal O}
(\epsilon^3 ) \nonumber \\ &=&
 \sqrt{g} + \epsilon \left[
{\partial \sqrt{\widetilde{g}} \over \partial \epsilon }
\right]_{\epsilon = 0}
+
{\epsilon^2 \over 2}\left[
{\partial^2 \sqrt{\widetilde{g}}
\over \partial \epsilon^2 }
\right]_{\epsilon = 0}
+ {\cal O} (\epsilon^3 )\,.
\end{eqnarray}
We  calculate
\begin{equation}
{\partial \sqrt{\widetilde{g}}
\over \partial \epsilon }
= {1\over 2} \sqrt{\widetilde{g}} \; \widetilde{g}^{ab} \; {\partial
\widetilde{g}_{ab}
\over \partial \epsilon}
= \sqrt{\widetilde{g}} \; [ \widetilde{g}^{ab}
({\bf e}_a \cdot {\bf W}_b )
+ \epsilon \widetilde{g}^{ab} ({\bf W}_a \cdot {\bf W}_b )]\,.
\label{eq:sqrt1e}
\end{equation}
Taking the $\epsilon \to 0$ limit,
we have
\begin{equation}
\sqrt{g}_{(1)} =
\epsilon \; \sqrt{g} \; ({\bf e}_a \cdot {\bf W}^a )\,.
\end{equation}
At second order, we have
\begin{eqnarray}
{\partial^2 \sqrt{\widetilde{g}}
\over \partial \epsilon^2 }
 &=&
\left({\partial \sqrt{\widetilde{g}} \over \partial \epsilon} \right)
[ \widetilde{g}^{ab} ({\bf e}_a \cdot {\bf W}_b ) + \epsilon ({\bf W}_a
\cdot {\bf W}_b )] +
\sqrt{\widetilde{g}} \; \widetilde{g}^{ab} ({\bf W}_a \cdot {\bf W}_b )
\nonumber \\
&+&
\sqrt{\widetilde{g}} \left({\partial \widetilde{g}^{ab} \over \partial
\epsilon} \right)
[({\bf e}_a \cdot {\bf W}_b )+ \epsilon ({\bf W}_a \cdot {\bf W}_b )]\,.
\end{eqnarray}
Using  (\ref{eq:gabi}) and (\ref{eq:sqrt1e}) in this expression, and
taking the $\epsilon \to 0$ limit gives the second order correction
in  (\ref{eq:sqrt}).

Finally we consider the intrinsic scalar curvature $\widetilde{R}$.
We  restrict our attention to the first order correction.
We use the Palatini identity for the first order correction of the
Christoffel symbols $\Gamma^c_{ab}$,
\begin{equation}
\Gamma_{ab}^c{}_{(1)} = {1 \over 2} g^{cd}
( \nabla_b g_{ad\, (1)}
+ \nabla_a g_{bd\, (1)}
- \nabla_d g_{ab\, (1)} )\,,
\end{equation}
to obtain, using  (\ref{eq:gab}),
\begin{eqnarray}
\Gamma_{ab}^c{}_{(1)} &=&
 \epsilon (
{\bf e}^c \cdot \nabla_a {\bf W}_b
+ {\bf W}^c \cdot \nabla_a {\bf e}_b )
\nonumber \\
&=&  \epsilon [
{\bf e}^c \cdot \nabla_a {\bf W}_b
- K_{ab} ( {\bf W}^c \cdot {\bf n} )]\,,
\label{eq:gamma1}
\end{eqnarray}
where we have used the Gauss-Weingarten equation (\ref{eq:gw1})
in the second line, and the fact that the covariant derivative is
torsionless, {\it i.e.} $\nabla_a {\bf W}_b = \nabla_b {\bf W}_a$.
We recall that $\Gamma^c_{ab}{}_{(1)}$,
unlike $\Gamma^c_{ab}$, transforms as a tensor under surface 
reparametrizations.
Using the definition of the Riemann tensor given by
(\ref{eq:riemann}), we have that at first order the Riemann
tensor is
\begin{equation}
{\cal R}^a{}_{bcd\, (1) } =  \nabla_c \Gamma_{db}^a{}_{(1)}
 - \nabla_d \Gamma_{cb}^a{}_{(1)}\,,
\end{equation}
so that, inserting  (\ref{eq:gamma1}), we have
\begin{eqnarray}
\fl
{\cal R}^a{}_{bcd\, (1) } =  \epsilon [
&-& {\cal R}^e{}_{bcd} ( {\bf e}^a \cdot {\bf W}_e )
+ {\cal R}_{cdb}{}^e ( {\bf e}_e \cdot {\bf W}^a )
+ K_{bc}  ( {\bf n} \cdot \nabla_d {\bf W}^a )
\nonumber \\ \fl
&-& K_{c}{}^a ( {\bf n} \cdot \nabla_d {\bf W}_b )
+ K_{d}{}^a ( {\bf n} \cdot \nabla_c {\bf W}_b )
- K_{db} ( {\bf n} \cdot \nabla_c {\bf W}^a ) ]\,,
\end{eqnarray}
where we have used the definition of the Riemann tensor and
both the Codazzi-Mainardi equations (\ref{eq:cm1}) and the
the Gauss-Codazzi equations (\ref{eq:g1}).
For the Ricci tensor this implies
\begin{eqnarray}
{\cal R}_{bd\, (1)}
&=&  \nabla_c \Gamma_{bd}^c{}_{(1)}
 - \nabla_d \Gamma_{cb}^c{}_{(1)}
\nonumber \\
&=&  \epsilon [
- ( {\cal R}^e{}_{bad } + {\cal R}^e{}_{dab} ) ({\bf e}^a \cdot {\bf W}_e )
+ K_{ab}  ({\bf n} \cdot \nabla_d {\bf W}^a )
\nonumber \\
&+ & K_a{}^a ({\bf n} \cdot \nabla_a {\bf W}_b )
-  K_{db} ({\bf n} \cdot \nabla_c {\bf W}^c )
- K ({\bf n} \cdot \nabla_b {\bf W}_d )\,,
\end{eqnarray}
These expressions are valid for a hypersurface of arbitrary dimension.
Restricting our attention to 2-dimensional surfaces, and exploiting
the fact that both the Riemann tensor and the Ricci tensor can be
expressed in terms of the scalar curvature ${\cal R}$, we have that
\begin{equation}
g^{ab} {\cal R}_{ab\, (1)} =  \epsilon [
2 (K^{ab} - K g^{ab}) ({\bf n} \cdot \nabla_a {\bf W}_b )
-   {\cal R} ({\bf e}_a \cdot {\bf W}^a ) ]
\,,
\end{equation}
For the scalar curvature we have then
\begin{eqnarray}
{\cal R}_{(1)} &=& g^{ab}{}_{(1)}
 {\cal R}_{ab} + g ^{ab} {\cal R}_{ab\, (1)}
\nonumber \\
&=& 2 \epsilon [ ( K^{ab} - K g^{ab} ) ({\bf n}
\cdot \nabla_a {\bf W}_b ) -  {\cal R} ({\bf e}^a \cdot {\bf W}_a )]\,.
\label{eq:r1}
\end{eqnarray}
Note that it depends on two derivatives of the deformation vector
${\bf W}$, and that it involves the extrinsic geometry of the surface.

In general, it is always possible to expand any geometrical
quantity $\widetilde{f}$ on the deformed surface
$\widetilde\Sigma$ in terms of ${\bf W}$. With $\widetilde{f} = f
+ f_{(1)} + f_{(2)}$, besides direct computation, we can obtain
the second order term $f_{(2)}$ by `deforming' the first order
term $f_{(1)}$, that is via the important identity
\begin{equation}
f_{(2)} = {1 \over 2} \; \widetilde{f}_{(1)}\,.
\label{eq:rela}
\end{equation}
Here $\widetilde{f}_{(1)}$ is to be understood
as the expansion to order $\epsilon$ of $f_{(1)}$, as we illustrate below.
This alternative approach is particularly useful
when we consider global geometric quantities associated
with $\widetilde\Sigma$ (see Sections 6,8).
Note that, in agreement with the identity  (\ref{eq:rela}),
we have, for example,
\begin{eqnarray}
\fl
\sqrt{g}_{(2)} =  {\epsilon \over 2} \widetilde{\sqrt{g}}_{(1)}
&=& {\epsilon \over 2} \left[ \sqrt{g}_{(1)} ({\bf e}^a \cdot {\bf W}_a )
+ \sqrt{g} g^{ab}_{(1)} ({\bf e}_a \cdot {\bf W}_b )
+ \sqrt{g} ({\bf W_a} \cdot {\bf W}^a ) \right]
\nonumber \\ \fl
&=& {\epsilon^2 \over 2} \sqrt{g} [
({\bf e}^a \cdot {\bf W}_a )^2
- ({\bf e}^a \cdot {\bf W}^b ) ({\bf e}_a \cdot {\bf W}_b )
- ({\bf e}^b \cdot {\bf W}^a )
({\bf e}_a \cdot {\bf W}_b )
\nonumber \\ \fl
&+& ({\bf W}_a \cdot {\bf W}^a )]\,,
\end{eqnarray}
which, using the completeness relation (\ref{eq:comple}) in the last term,
reproduces the second order contribution to (\ref{eq:sqrt}).
The identity (\ref{eq:rela}) can be proved using variational 
techniques. It does not appear to be available in the literature.
Of course, this could be just a shortcoming of our search.

\section{ Deformation of the extrinsic geometry}

Let us turn now to the extrinsic geometry of the deformed
surface $\widetilde\Sigma$.
For its unit normal $\widetilde{\bf n}$, we use the defining relations
$\widetilde{\bf n} \cdot \widetilde{\bf e}_a = 0$, $\widetilde{\bf n}
\cdot \widetilde{\bf n} = 1$,
together with  (\ref{eq:tan}). We expand $\widetilde{\bf n}$
and we obtain the relations
\begin{eqnarray}
\epsilon ({\bf n} \cdot {\bf W}_a ) + ({\bf n}_{(1)} \cdot {\bf e}_a )
&=& 0\,,
\nonumber \\
\epsilon ( {\bf n}_{(1)} \cdot {\bf W}_a ) + ( {\bf n}_{(2)} \cdot
{\bf e}_a )
&=& 0\,,
\nonumber \\
{\bf n} \cdot {\bf n}_{(1)} &=& 0\,,
\nonumber \\
{\bf n}_{(1)}\cdot {\bf n}_{(1)} + 2 {\bf n} \cdot {\bf n}_{(2)}
&=& 0\,,
\nonumber
\end{eqnarray}
which provide six equations for the six unknowns ${\bf n}_{(1)}, {\bf n}_{(2)}$.
Some simple algebra gives
\begin{equation}
\fl
\widetilde{\bf n} = {\bf n}
- \epsilon ({\bf n} \cdot  {\bf W}_a ) {\bf e}^a
+ \epsilon^2 [ ({\bf n} \cdot  {\bf W}_b )
({\bf e}^b \cdot  {\bf W}^a ) {\bf e}_a
- {1 \over 2}
({\bf n} \cdot  {\bf W}^a )({\bf n} \cdot {\bf W}_a )
{\bf n}]  + {\cal O} (\epsilon^3 )\,.
\label{eq:not}
\end{equation}
Note that if the deformation is such that $ ({\bf n} \cdot  {\bf W}_a ) =
0$ the normals to the two surfaces coincide. This happens
for parallel surfaces, defined by ${\bf W} = a {\bf n}$, with $a$ constant
\cite{Struik}.

The extrinsic curvature of $\widetilde\Sigma$ is
$
\widetilde{K}_{ab} := - \widetilde{\bf n} \cdot \partial_a \widetilde{\bf
e}_b\,.
$
Expanding the right hand side to second order in $
\epsilon$, we obtain
\begin{equation}
\fl
\widetilde{K}_{ab} =
K_{ab} - {\bf n}_{(1)} \cdot \partial_a
{\bf e}_b - {\bf n} \cdot \partial_a {\bf e}_{b\,(1)}
- {\bf n}_{(1)} \cdot
\partial_a {\bf e}_{(1)}
- {\bf n}_{(2)} \cdot \partial_a {\bf e}_{b\, (1)}
 + {\cal O} (\epsilon^3 )
\,.
\end{equation}
We use  (\ref{eq:tan}), (\ref{eq:not}), together with
the Gauss-Weingarten equation for $\Sigma$,  (\ref{eq:gw1}).
We obtain
\begin{equation}
\fl
\widetilde{K}_{ab} = K_{ab} - \epsilon ( {\bf n} \cdot  \nabla_a {\bf
W}_{b} )
+ \epsilon^2 [ ( {\bf n} \cdot {\bf W}_c )
( {\bf e}^c \cdot \nabla_a {\bf W}_b )
- {1 \over 2} K_{ab} ( {\bf n} \cdot {\bf W}_c )
( {\bf n} \cdot {\bf W}^c ) ] + {\cal O} (\epsilon^3 )
\,.
\label{eq:kab}
\end{equation}
We note that $\widetilde{K}_{ab}$ transforms covariantly under
reparametrizations of $\Sigma$ and that it involves
two derivatives of the deformation vector ${\bf W}$.
Note that for parallel surfaces, one has that
$K_{ab\, (1)} = \epsilon \, a K_a{}^c K_{cb}$
and $K_{ab\, (2)} = 0$.

For the trace of the extrinsic curvature $K$,
Eqs (\ref{eq:gabi}), (\ref{eq:kab}) imply
\begin{eqnarray}
\fl
\widetilde{K} = K &-& \epsilon [
( {\bf n} \cdot \nabla_a {\bf W}^a )
+ 2  K^{ab}
( {\bf e}_a \cdot {\bf W}_b )]
\nonumber \\ \fl
&+& \epsilon^2 [ 2 ( {\bf e}^a \cdot {\bf W}^b )
( {\bf n} \cdot \nabla_a {\bf W}_b) +
2 K_{ab} ( {\bf e}^a \cdot {\bf W}_c )( {\bf e}^c \cdot {\bf W}^b )
\nonumber \\ \fl
&+& K_{ab}
( {\bf e}^a \cdot {\bf W}_c )( {\bf e}^b \cdot {\bf W}^c )
+  ( {\bf n} \cdot {\bf W}_c )( {\bf e}^c  \cdot \nabla_a{\bf W}^a )
\nonumber \\ \fl
&-& K_{ab} ( {\bf n} \cdot {\bf W}^a )( {\bf n} \cdot {\bf W}^b )
- {K \over 2} ( {\bf n} \cdot {\bf W}^c )( {\bf n} \cdot {\bf W}_c ) ]
 + {\cal O} (\epsilon^3 )
\,.
\label{eq:k}
\end{eqnarray}

At this point we can use the Gauss-Codazzi equation (\ref{eq:g3})
to check the validity of these expressions for the deformation
of the extrinsic curvature.
At first order, using  (\ref{eq:gabi}),
(\ref{eq:kab}),
(\ref{eq:k}),
we reproduce  (\ref{eq:r1}).
Note that the check is non trivial; it requires
extensive use, in various degrees of contraction,
of the Gauss-Codazzi-Mainardi equations.

\section{ Deformations of the geometry: decomposed}

The existence of a basis adapted to the surface $\Sigma$,
$ \{ {\bf e}_a , {\bf n} \} $, suggests a natural
decomposition of the deformation vector ${\bf W}$ into its
tangential and normal components
(see {\it e.g.} \cite{Defo}),
\begin{equation}
{\bf W} = \Phi^a {\bf e}_a + \Phi {\bf n}\,,
\end{equation}
It follows that the projections of the first two derivatives of the
deformation vector are
\begin{eqnarray}
{\bf e}_a \cdot {\bf W}_b &=& \nabla_b \Phi_a + \Phi K_{ab}\,,
\label{eq:c1}\\
{\bf n} \cdot {\bf W}_a &=& \nabla_a \Phi - \Phi^b K_{ab}\,,
\label{eq:c2}\\
{\bf e}_c \cdot \nabla_a {\bf W}_b &=& \nabla_a \nabla_b \Phi_c
- \Phi^d K_{bd} K_{ac} + 2 K_{c(a} \nabla_{b)}  \Phi
+ \Phi \nabla_a K_{bc}\,,
\label{eq:c3}\\
{\bf n} \cdot \nabla_a {\bf W}_b &=& \nabla_a \nabla_b \Phi
- \Phi K_{ac} K^c{}_b - 2 K_{c(a} \nabla_{b)} \Phi^c
- \Phi^c \nabla_c K_{ab}\,,
\label{eq:c4}
\end{eqnarray}
where we have used the Gauss-Weingarten equations (\ref{eq:gw1}),
(\ref{eq:gw2}), and the contracted Codazzi-Mainardi equation (\ref{eq:cm2}).

Using these expressions, the basic geometric quantities that
characterize $\widetilde\Sigma$
take the form, to first order in $\epsilon$,
\begin{eqnarray}
g_{ab\, (1)} &=& \epsilon \left( 2 K_{ab} \Phi + 2 \nabla_{(a} \Phi_{b)}\right)\,,
\label{eq:gab1}
\\
\sqrt{g}_{(1)} &=& \epsilon \sqrt{g} \; ( K \Phi + \nabla_a \Phi^a )\,,
\label{eq:sqrtg1}
\\
K_{ab\, (1)} &=& \epsilon \left( - \nabla_a \nabla_b
\Phi + K_{ac} K^c{}_b \Phi +
\Phi^c \nabla_c K_{ab} + 2 K_{c(a} \nabla_{b)} \Phi^c
\right)\,,\label{eq:kt}
\\
K_{(1)} &=& \epsilon \left( - \nabla^2 \Phi - K_{ab} K^{ab} \Phi
+ \Phi^c \nabla_c K \right)\,,
\\
{\cal R}_{(1)} &=& \epsilon \left[ 2 (K^{ab} - K g^{ab}) \nabla_a \nabla_b \Phi
+ \Phi^c \nabla_c {\cal R} \right]\,,
\end{eqnarray}
We can thus identify both a normal (linear in $\Phi$), and a tangential
deformation (linear in $\Phi^a$) of these geometrical quantities.
We note that these expressions coincide with
the ones obtained  {\it e.g.} in \cite{Defo}.

In particular, to first order
the tangential components correspond to an
infinitesimal (active) reparametrization of the surface.
Indeed, each of
the three surface scalars ${\bf X} (\xi^a )$ transforms as
\begin{equation}
\delta_{\mbox {\sf rep}} {\bf X} = v^a (\xi^b) \,\partial_a {\bf X} =
v^a {\bf e}_a\,,
\end{equation}
under a reparametrization $\xi^a\to\xi^a -v^a$.
This is exactly the effect of a tangential deformation at first order
with the identification of $v^a$
with the surface vector field
defined by the projection $\Phi^a ={\bf W}\cdot
{\bf e}^a$. Reflecting this fact, at first order,
geometrical quantities transform as surface Lie derivative along
the surface vector field $\Phi^a$. For example, setting $\Phi =
0$ in  (\ref{eq:gab1}), we have that
\begin{equation}
g_{ab\, (1) \mbox{\sf tang.}} =
2 \nabla_{(a} \Phi_{b)} = {\cal L}_{\Phi^a} g_{ab}\,,
\end{equation}
where ${\cal L}_{\Phi^a}$ denotes the surface Lie derivative
along $\Phi^a$.

It is important to emphasize that,
having made this identification
at first order, the tangential part of any
total geometrical invariant of $\widetilde\Sigma$ is always given
by the  integral of a total divergence, which vanishes over a
closed surface without boundaries. To see this, consider an
invariant $I = \int dA f (\xi^a)$, where $f (\xi^a )$ is a scalar
under reparametrizations. Since $f$ is a scalar we have simply
that
\begin{equation}
f_{(1) \mbox{\sf tang.}} = \Phi^a \partial_a f \,,
\end{equation}
no matter how complicated the dependence of $f$ on the geometry might be.
Moreover, setting $\Phi =0$ in   (\ref{eq:sqrtg1}),  we
have $
\sqrt{g}_{(1)\mbox{\sf tang.} } = \sqrt{g} \; \nabla_a \Phi^a $,
and therefore
\begin{equation}
I_{(1) \mbox{\sf tang.}} = \int dA \nabla_a ( \Phi^a f )\,.
\end{equation}

At first order, we can always disentangle the physical normal
deformation  from reparametrizations and we can safely set
$\Phi^a$ to vanish. Matters, however, are not so simple at second
order. Let us consider the second order variation of the metric.
Using the completeness relation (\ref{eq:comple}) it takes the
form
\begin{eqnarray}
\fl
g_{ab\,(2)} &=& \epsilon^2 [
({\bf e}_a \cdot {\bf W}_b )({\bf e}^a \cdot {\bf W}^b )+
({\bf n} \cdot {\bf W}_a)({\bf n} \cdot {\bf W}^a)] \nonumber\\ \fl
&=& \epsilon^2 [ (\nabla_b \Phi_a + \Phi K_{ab})(\nabla^b \Phi^a + \Phi K^{ab})
+ (\nabla_a \Phi - \Phi^c K_{ac})(\nabla^a\Phi - \Phi_d K^{ad})]
\nonumber\\ \fl
 &=& \epsilon^2 [ \nabla_a \Phi \nabla_b\Phi +
K_{a}{}^c K_{cb} \,\Phi^2 +
2 K_{c(a} \nabla_{b)} ( \Phi \Phi^c ) \nonumber \\ \fl
&+& \nabla_a \Phi^c \nabla_b \Phi_c + K_{ac} K_{bd}\,\Phi^c \Phi^d ]\,,
\label{eq:d2g}
\end{eqnarray}
and with (\ref{eq:gab1}) this completely describes the
deformation of the metric to all orders. At second order, normal
and tangential deformations begin to talk to each other.
When both normal and tangential deformations are present, there
is a mixing. The purely tangential deformation at this order is
certainly not simply a second order reparametrization, in the
sense of a composition of Lie derivatives. This might appear to
be obvious: the second order tangential deformation involves the
extrinsic geometry through the quadratic term in $K_{ab}$,
whereas a reparametrization is a purely intrinsic concept and as
such it should not involve $K_{ab}$. One has to be careful,
however, to check how the dependence on the extrinsic geometry enters.

Let us look more closely at the issue of reparametrization
covariance at second order. Consider an infinitesimal change of coordinates
on the surface $\xi^a \to \xi'^a = \xi^a + v^a $, and a surface scalar field
$ f (\xi^a )$. By definition of scalar field we have
$ f' (\xi'^a ) = f (\xi^a)$. In order to evaluate the
change in the scalar field at the same point, we
expand $f' (\xi'^a ) = f' (\xi^a + v^a )$ in powers
of $v^a$,
\begin{equation}
f' (\xi'^a ) =
f' (\xi^a ) +
 v^a \partial_a f' (\xi^a )
+ {1 \over 2} (v^a  \partial_a )(v^b \partial_b) f (\xi^a
) + \cdots\,,
\end{equation}
where we replace $f'$ by $f$ in the last term since it is already
of second order in $v^a$. For the middle term note that
\begin{eqnarray}
 v^a \partial_a f' (\xi^a )
&=& v^a \partial_a [ f' (\xi'^a - v^a )] \nonumber \\
&=&  v^a \partial_a [ f' (\xi'^a ) - v^b \partial_b
f (\xi^a) ] \nonumber \\
&=&  v^a \partial_a [ f(\xi^a ) - v^b \partial_b
f (\xi^a) ]\,.
\end{eqnarray}
Therefore, to second order, we have
\begin{equation}
\fl
\delta_{\mbox{rep}} f (\xi^a )
= f' (\xi^a ) - f (\xi^a )
= - v^a \partial_a f (\xi^a )
+ {1 \over 2} (v^a  \partial_a)(v^b  \partial_b)
f (\xi^a )\,.
\end{equation}
At first order, we have minus the Lie derivative of the
scalar field, since now we are considering passive
transformations. The second order contribution is
the composition of two Lie derivatives.

In particular, for the embedding functions
we obtain
\begin{equation}
\delta_{\mbox{rep}} {\bf X}
= - v^a  {\bf e}_a
+ {1 \over 2} \left( v^a \nabla_a v^b {\bf e}_b -
v^a v^b  K_{ab} {\bf n} \right)\,.
\label{eq:delrepX}
\end{equation}
The important point here is that, at second order, a
reparametrization will generally alter the embedding functions.
In contrast, by construction ${\bf X}$ is only modified at first
order in ${\bf W}$. Note also that, at second order, a
reparametrization produces a change in ${\bf X}$ along the {\it
normal} direction. Moreover, it depends explicitly on the
extrinsic geometry. This justifies our earlier caveat.

The tangent vectors ${\bf e}_a$ transform as covariant surface vectors under
reparametrization: we have at first order for a vector $f_a$
\begin{equation}
\delta_{\mbox{rep}} f_{a\, (1)}
= - v^b \partial_b f_a -  \partial_a v^b\, f_b\,.
\end{equation}
Thus for ${\bf e}_{a}$:
\begin{eqnarray}
\delta_{\mbox{rep}} {\bf e}_{a\, (1)}
&=& - v^b \partial_b {\bf e}_a -  \partial_a v^b\, {\bf e}_b
\nonumber\\
&=& - v^b \nabla_b {\bf e}_a - \nabla_a v^b\, {\bf e}_b \nonumber\\
&=&  v^b K_{ab} {\bf n}  -  \nabla_a v^b \, {\bf e}_b\,,
\end{eqnarray}
involving both tangential and normal parts. Note how the normal contribution projects out
of $\delta_{\mbox{rep}} g_{ab\,(1)} =
2{\bf e}_{(a}\cdot  \delta_{\mbox{rep}}
{\bf e}_{b)\,(1)}= 2\nabla_{(a} v_{b)}$.
At second order,
\begin{equation}
\delta_{\mbox{rep}} {\bf e}_{a\,(2)}
= {1\over 2}\left( - v^b \partial_b \delta_{\mbox{rep}} {\bf e}_{a\,(1)}
- \delta_{\mbox{rep}} {\bf e}_{b\,(1)} \partial_a v^b \right)\,.
\end{equation}
Similarly, we obtain for the metric at second order
 \begin{eqnarray}
\delta_{\mbox{rep}} g_{ab\,(2)} &=& {1\over 2}(
v^c (\nabla_c \nabla_a v_b + \nabla_c\nabla_b v_a)
+ (\nabla_c v_a) (\nabla_b v^c) 
\nonumber \\ 
&+&
(\nabla_c v_b) (\nabla_a v^c) +
2 (\nabla_a v_c)(\nabla_b v^c)\,.
\end{eqnarray}
Note that $\delta_{\mbox{rep}} g_{ab\,(2)}$ is manifestly intrinsic
and, as such, distinct from the contribution to $g_{ab\,(2)}$ quadratic in $\Phi^a$
on setting  $\Phi^a= v^a$.
We note also that
this expression coincides with the metric at second order induced by 
(\ref{eq:delrepX})
\begin{equation}
 \delta_{\mbox{rep}} g_{ab\, (2)} = 2{\bf e}_{(a}\cdot  \delta_{\mbox{rep}}
{\bf e}_{b)\,(2)}
+ \delta_{\mbox{rep}} {\bf e}_{a\,(1)}\cdot
\delta_{\mbox{rep}} {\bf e}_{b\,(1)}\,.
\end{equation}
At second order,  we see that we cannot disentangle
the physical normal deformation  from reparametrizations
and we cannot set $\Phi^a$ to vanish. However,
as we will see below in Sect. 8, when considering the second
order deformation of global geometric invariants,
the tangential component of the deformation will appear
only in boundary terms or in terms that vanish when the
membrane is at equilibrium.

\section{ Variation of the Helfrich-Canham Hamiltonian: first order}

Let us expand the Hamiltonian $F[{\bf X}]$ as given by (\ref{eq:model})
to first order in $\epsilon$.
This will allow us to identify its Euler-Lagrange
derivative, and the equilibrium conditions for the vesicle.
(For a different approach emphasizing the normal component
of the deformation, see \cite{Defo}.)

The expansion  can always be written in the form
\begin{equation}
F_{(1)} [{\bf X}, {\bf W}] = \epsilon \int dA \; {\bf  E}_F \cdot {\bf W}
+ \epsilon \int dA \; \nabla_a {\cal Q}^a\,.
\label{eq:F1}
\end{equation}
Here ${\bf  E}_F$ denotes the Euler-Lagrange derivative for $F
[{\bf X}]$. The quantity ${\cal Q}^a$ appearing in the total
divergence in the second term is the Noether current
\cite{Stress}, which will be used below in Sect. 7  to derive the
stresses and torques acting on the surface associated with $F
[{\bf X}]$.

We use the results of Sect. 2 to derive, term by term,
the various contributions to (\ref{eq:F1}). For the area
of the vesicle, we find using (\ref{eq:sqrt}) that
\begin{equation}
A_{(1)} = \epsilon \int dA \; ( {\bf e}_a \cdot {\bf W}^a )\,.
\end{equation}
To cast this expression in the form (\ref{eq:F1}), we integrate by
parts, and obtain
\begin{equation}
A_{(1)} = \epsilon \int dA \; K {\bf n} \cdot {\bf W}
+ \epsilon \int dA \; \nabla_a ( {\bf e}^a \cdot {\bf W})\,.
\label{eq:a11}
\end{equation}
Therefore, the Euler-Lagrange derivative
of the area is purely normal, and proportional
to the mean extrinsic curvature, ${\bf E}_A = E_A {\bf n} = K {\bf n}$.
The first feature is common to all reparametrization invariants:
to first order in $\epsilon$, tangential deformations contribute
only boundary terms, as shown in the previous section.
The latter tells us that minimal surfaces, extremizing the area,
have vanishing mean extrinsic curvature, $ E_A = K= 0$.

Note that for a constant normal displacement ${\bf W} = a {\bf n}$ we have
\begin{equation}
A_{(1)} = a \int dA \; K = a M\,,
\end{equation}
so that the total mean extrinsic curvature is proportional to the
area difference in the normal direction.

If we require that the area be infinitesimally locally invariant,
then we have a constraint on ${\bf W}$ of the form $ {\bf e}_a
\cdot {\bf W}^a=0$. This does not, however, alter the value of
the Euler-Lagrange derivative.

For the volume enclosed by the vesicle, we use the definition
(\ref{eq:vol}), together with (\ref{eq:sqrt}), (\ref{eq:not}), to
derive
\begin{eqnarray}
V_{(1)} &=& {\epsilon \over 3} \left[ \int dA_{(1)} \; ({\bf n}
\cdot {\bf X}) + \int dA \; \left( {\bf n}_{(1)} \cdot {\bf X}
+ \epsilon {\bf n} \cdot {\bf W} \right) \right] \nonumber \\
&=&  {\epsilon \over 3} \int dA \; [ ({\bf W} \cdot {\bf n})
+ ({\bf e}_a \cdot {\bf W}^a ) ({\bf n} \cdot {\bf X} )
- ({\bf n} \cdot {\bf W}^a ) ({\bf e}_a \cdot {\bf X}) ]
\nonumber
\end{eqnarray}
We integrate by parts the second and third terms and neglect a
total divergence to obtain
\begin{equation}
V_{(1)} =  \epsilon \int dA \; {\bf n} \cdot {\bf W}\,,
\label{eq:vol1}
\end{equation}
therefore we find that the Euler-Lagrange derivative of the
enclosed volume functional is simply unity, $E_V = 1$.

Let us consider now the total mean extrinsic curvature, $M$, as
defined in (\ref{eq:total}). We use (\ref{eq:sqrt}) and
(\ref{eq:k}) to derive
\[
M_{(1)} = \epsilon \int dA \; [ K ({\bf e}^a \cdot {\bf W}_a )
- 2 K^{ab} ({\bf e}_a \cdot {\bf W}_b )
- ({\bf n} \cdot \nabla_a {\bf W}^a )]\,.
\]
To put it in the form (\ref{eq:F1}), we integrate by parts and use
the Gauss-Weingarten equations (\ref{eq:gw1}), (\ref{eq:gw2}) to
arrive at
\begin{equation}
\fl
M_{(1)} = \epsilon \int dA \; {\cal R} \; ({\bf n} \cdot {\bf W})
+ \epsilon \int dA \; \nabla_a [ (K g^{ab} - K^{ab} ) ({\bf e}_b
\cdot {\bf W}) - ({\bf n} \cdot {\bf W}^a )]\,.
\label{eq:m11}
\end{equation}
The scalar intrinsic curvature appears as the Euler-Lagrange derivative
of the total mean extrinsic curvature functional $E_M = {\cal R}$.

As mentioned above, the Gaussian bending energy $F_G$ as given by
(\ref{eq:gaussian}) is a topological invariant. As such, we
expect both $F_{G \, (1)}$ and $F_{G\, (2)}$ to be a total
divergence. As this provides a non-trivial check, let us consider
its expansion to first order. Moreover, in any case, we are
interested in the non-vanishing contribution to the Noether
charge. Using  (\ref{eq:area}), (\ref{eq:r1}), we have
\begin{equation}
F_{G \, (1)}
= \epsilon  \alpha_G \int dA \; [ 2  ( K^{ab} - K g^{ab} ) ({\bf n}
\cdot \nabla_a {\bf W}_b ) -  {\cal R} ({\bf e}^a \cdot {\bf W}_a )]\,.
\end{equation}
We integrate the first two terms by parts, and use the
the second Gauss-Weingarten equation (\ref{eq:gw2}),
together with the contracted Codazzi-Mainardi equations (\ref{eq:cm2}), to obtain
\begin{eqnarray}
F_{G\, (1)} &=& \epsilon \alpha_G \int dA \; [ - {\cal R} g^{ab}
+ 2 K K^{ab} - 2 K^{ac} K^b{}_c ] ({\bf e}_a \cdot {\bf W}_b )
\nonumber \\
&+& \epsilon \alpha_G \int dA \nabla_a [
2 ( K^{ab} - K g^{ab} ) ({\bf n} \cdot {\bf W}_b )
- {\cal R} ({\bf e}^a \cdot {\bf W} ) ]\,,
\label{eq:gb1}
\end{eqnarray}
where the first line vanishes because of  the contracted
Gauss-Codazzi equation (\ref{eq:g2}). Therefore $F_{G\, (1)}$ is
given by a total divergence. With the help of the relationship
(\ref{eq:rela}), the second order term $F_{G\, (2)}$ is a total
divergence as well. To see this, recall the fact that the
'deformation' of the divergence of a vector density is equal to
the divergence of the deformation. In this particular  case we
have
\begin{equation}
F_{G\,(1)} = \epsilon \alpha_G \int dA \; \nabla_a {\cal Q}^a_{G}
= \epsilon \alpha_G \int d^2 \xi \; \partial_a ( \sqrt{g} \; {\cal Q}^a_{G} )\,,
\end{equation}
where ${\cal Q}^a_G$ denotes the contribution of the Gaussian
bending rigidity to the Noether current, and it is given
explicitly by the argument of the covariant derivative in 
(\ref{eq:gb1}). At second order we have then the total divergence
\begin{equation}
F_{G\,(2)} = {1 \over 2} \widetilde{F}_{G\, (1)}
= {\epsilon \over 2} \alpha_G \int d^2 \xi \;
\partial_a [ ( \sqrt{g} \; {\cal Q}^a_{G} )_{(1)} ]\,.
\end{equation}

Finally, for the bending energy (\ref{eq:bending}), we obtain, using
 (\ref{eq:sqrt}), (\ref{eq:k}),
\begin{equation}
\fl
F_{b\, (1)} =
\epsilon \int dA \; [ K^2 ({\bf e}_a \cdot {\bf W}^a )
- 4 K K^{ab} ({\bf e}_a \cdot {\bf W}_b )
- 2 K ({\bf n} \cdot \nabla_a {\bf W}^a )]\,,
\end{equation}
and  integration by parts twice gives
\begin{eqnarray}
\fl
F_{b\, (1)} &=&
\epsilon \int dA \; \left[ - 2 \nabla^2 K + K^3 - 2 K K_{ab} K^{ab}
\right] {\bf n} \cdot {\bf W}
\nonumber \\ \fl
&+& \epsilon \int dA \; \nabla_a \left[ ( K^2 g^{ab} - 2 K K^{ab} )
({\bf e}_b \cdot {\bf W} )
+ 2 (\nabla^a K )({\bf n} \cdot {\bf W} )
- 2 K ({\bf n} \cdot {\bf W}^a )\right]\,.
\label{eq:b11}
\end{eqnarray}
Therefore the Euler-Lagrange derivative of the bending energy functional is
\begin{eqnarray}
E_{F_b} &=& - 2 \nabla^2 K + K^3 - 2 K K_{ab} K^{ab}
\nonumber \\
&=& - 2 \nabla^2 K + K (2 {\cal R} - K^2 )\,,
\end{eqnarray}
where we have used the Gauss-Codazzi equation (\ref{eq:g3}) to
obtain the second line.

We are now in a position to write down the equilibrium
conditions for the Hamiltonian (\ref{eq:model}). We set
\begin{equation}
{\bf E}_F = E_F \; {\bf n}\,,
\end{equation}
where
\begin{equation}
E_F = \alpha \; \left[ - 2 \nabla^2 K + K (2 {\cal R} - K^2 ) \right]
+ \mu K + \beta {\cal R} - P \,.
\label{eq:EL}
\end{equation}
Then equilibrium configurations that extremize the Helfrich-Canham Hamiltonian
satisfy
\begin{equation}
E_F = 0\,.
\label{eq:shape}
\end{equation}
This is known as the shape equation \cite{Hel.OuY:87}. Note that
it is a nonlinear 4th order partial differential equation.
Progress in the understanding of its space of solutions has been
limited to the special case of axisymmetric
configurations, see {\it e.g.}  \cite{Seifert}.

\section{ Stresses and torques}

The Noether current appearing in (\ref{eq:F1}) allows the
derivation of the stresses and the torques acting on the
membrane, using the invariance under rigid motions in space.
This was done in \cite{Stress} by decomposing the
deformation in its normal and tangential parts. Here, we
provide an alternative derivation, which has the advantage
of being more direct. See also
\cite{VirgaRosso,RSV}.

Before, we proceed, for the purposes of this section it is
convenient to rewrite the Helfrich-Canham Hamiltonian
(\ref{eq:model}) by isolating the volume term as
\begin{equation}
F = F_{s} - P\; V\,,
\end{equation}
with the surface part of the Hamiltonian $F_s =
F_b + F_G + \mu A + \beta M$.

Collecting the various total surface divergences appearing in
 (\ref{eq:a11}), (\ref{eq:m11}), (\ref{eq:gb1}),
(\ref{eq:b11}), it is straightforward to identify the Noether
current associated with $F_s$,
\begin{eqnarray}
\fl
{\cal Q}^a &=& \alpha \; \left[ ( K^2 g^{ab} - 2 K K^{ab} ) ({\bf e}_b
\cdot {\bf W})
+ 2 (\nabla^a K )({\bf n} \cdot {\bf W} )
- 2 K ({\bf n} \cdot {\bf W}^a )\right] \nonumber \\ \fl
&+& \beta [ (K g^{ab} - K^{ab} ) ({\bf e}_b
\cdot {\bf W}) - {\bf n} \cdot {\bf W}^a ] + \mu \; ( {\bf e}^a
\cdot {\bf W})
\nonumber \\ \fl
&+& 2 \alpha_G 
( K^{ab} - K g^{ab} ) ({\bf n} \cdot {\bf W}_b )\,.
\label{eq:noether}
\end{eqnarray}
Note that although the Gaussian bending energy does not contribute
to the Euler-Lagrange equations, it does appear in the Noether current
${\cal Q}^a$.

We consider now a (simply connected) piece of the membrane, which
we denote by $\Sigma_0$, bounded by a curve $C$, and we specialize
the variation of the Hamiltonian (\ref{eq:F1}) to this arbitrary
region of the membrane. We have that
\begin{equation}
F_{s\,(1)} = \epsilon \int_{\Sigma_0}  dA \; \left[ \; E_{F_s} \; {\bf  n}
\cdot {\bf W}
+  \; \nabla_a {\cal Q}^a \right]\,.
\label{eq:F10}
\end{equation}
We exploit the invariance of the Hamiltonian under rigid motions
in space. First, we consider an infinitesimal translation ${\bf
W}={\bf a}$, with ${\bf a}$ constant. As the Hamiltonian is
invariant under translations, the left hand side of (\ref{eq:F10})
vanishes, and with the stress tensor ${\bf f}^a$ defined by
\begin{equation}
{\cal Q}^a = - {\bf a} \cdot {\bf f}^a \,,
\end{equation}
it follows that
we can write the Euler-Lagrange derivative as a conservation law
\begin{equation}
E_{F_s} \; {\bf n} = \nabla_a {\bf f}^a\,,
\end{equation}
where the stresses associated with the Hamiltonian
(\ref{eq:model}) are given by
\begin{equation}
\fl
{\bf f}^a = - \alpha \; \left[ ( K^2 g^{ab} - 2 K K^{ab} ){\bf e}_b
+ 2 (\nabla^a K ){\bf n} \right] - \beta (K g^{ab} - K^{ab} )
{\bf e}_b  - \mu \;  {\bf e}^a \,.
\end{equation}
We emphasize that it is far from obvious from
the shape equation itself (\ref{eq:shape}) that it can be written
as a conservation law.

There are three conservation laws, and only one shape equation.
This is a consequence of the reparametrization invariance of the
Hamiltonian. This statement  can be made explicit using the decomposition
of the stress tensor into tangential and normal parts as follows:
\begin{equation}
{\bf f}^a = f^{ab} \; {\bf e}_b + f^a \; {\bf n}\,.
\end{equation}
The surface covariant derivative then gives
\begin{eqnarray}
\nabla_a f^a - K_{ab} f^{ab} &=& E_{F_s} = P \,,
\label{eq:dec1}
\\
\nabla_a f^{ab} + K^{b}{}_a f^a &=& 0\,.
\label{eq:dec2}
\end{eqnarray}
The first equation is the shape equation expressed in terms of the
projections $f^a$ and $f^{ab}$. The second equation expresses
the content of reparametrization invariance as a consistency
check: the normal stress and the tangential stress must
balance exactly in this way. Note that this identity is
 potentially useful in numerical simulations, where reparametrization
invariance is necessarily lost, and one is interested
in quantifying the degree of violation.

The physical meaning of the stress tensor ${\bf f}^a $ is perhaps
best illustrated by considering the total force per unit length
acting on the curve $C$. Concretely, $C$ may be the shape of
an edge of the membrane \cite{Edge}, or the line boundary
between the two phases of a two-component vesicle \cite{JL}.
If we consider a basis $\{ {\bf t}, {\bf l} \} $ on the surface
adapted to the curve $C$ that bounds $\Sigma_0$, with ${\bf t}$
tangent to $C$, and ${\bf l} = l^a {\bf e}_a  $ the (outward) normal
to $C$ on the surface,
we  obtain the force per unit length acting on $C$,
$l_a {\bf f}^a = {\bf f}$, as
\begin{equation}
\fl
{\bf f} = \left[ K_{\parallel \perp} ( 2 \alpha K + \beta ) \right] {\bf
t}
+ \left[ \alpha K (K_\perp - K_\parallel ) 
- \beta K_\parallel - \mu\right] {\bf l}    - 2 \alpha (\nabla_\perp K ) {\bf
n}\,,
\end{equation}
where we denote the projections of the extrinsic curvature onto
the surface as $K_\parallel = K_{ab} t^a t^b$,  $K_\perp = K_{ab}
l^a l^b $, and $K_{\perp\parallel} = K_{ab} t^a l^b$. Note that
$K = K_\perp + K_\parallel$, and ${\cal R} = 2 (K_\parallel
K_\perp - K_{\perp\parallel}^2 )$. $\nabla_\perp = l^a
\nabla_\perp$ denotes the covariant derivative along the
direction normal to the curve $C$.

Similarly as we showed for translations, for an infinitesimal
rotation of the form ${\bf W}= {\bf b}\times {\bf X}$, we can
obtain the torques acting on the surface associated with the
Hamiltonian (\ref{eq:model}). We define the total angular momentum
${\bf m}^a$
\begin{equation}
{\cal Q}^a = - {\bf b} \cdot {\bf m}^a\,,
\end{equation}
where the torque ${\bf m}^a$ can be split in its
'orbital' and 'differential' parts as
\begin{equation}
{\bf m}^a = {\bf X} \times {\bf f}^a + {\bf s}^a\,.
\end{equation}
From the Noether charge (\ref{eq:noether}) we obtain directly that the
terms contributing to ${\bf s}^a$ are the ones involving derivatives
of the deformation vector, ${\bf W}_a$, so that
\begin{equation}
{\bf s}^a = [( 2 \alpha K + \beta )g^{ab} + 2 \alpha_G
(K g^{ab} - K^{ab} )   ] {\bf
e}_b \times {\bf
n}\,.
\end{equation}
Note that it is tangential to the surface.

The differential torque and the stress tensor are related by \cite{Stress}
\begin{equation}
\nabla_a {\bf s}^a = {\bf f}^a \times {\bf e}_a\,.
\end{equation}
We emphasize that this expression is valid also when
not in equilibrium.

\section{ Variation of the Helfrich-Canham Hamiltonian: second order}

In this section, we exploit the general results of Sects. 2, 3 to
derive the expansion to second order in $\epsilon$ of the
Hamiltonian (\ref{eq:model}). As we did at first order, we derive
the various terms that contribute to it. There are two possible
strategies: on one hand we can perform a direct expansion,
alternatively we can exploit the identity (\ref{eq:rela}), and
deform the first order terms we have obtained in Sect. 6. We will
adopt the most convenient strategy for each term. This part is a
straightforward calculation. However, both in order to have a
better understanding of the final result and to make contact with
the variational approach of  \cite{Defo}, we decompose the
deformation vector ${\bf W}$ in components. This is a
straightforward calculation as well, but it turns out that it is
possible to organize the result in a way which isolates boundary
terms and a contribution proportional to the Euler-Lagrange
derivative of each term.

For the area, we have immediately, using  (\ref{eq:sqrt}), that
\begin{equation}
\fl
A_{(2)} = {\epsilon^2 \over 2} \int dA
\; [ ( {\bf n} \cdot {\bf W}^a )( {\bf n} \cdot {\bf W}_a )
+ ( {\bf e}_a \cdot {\bf W}^a )( {\bf e}_b \cdot {\bf W}^b )
- ( {\bf e}_b \cdot {\bf W}^a )( {\bf e}_a \cdot {\bf W}^b ) ]\,.
\label{eq:area}
\end{equation}

The direct expansion at second order of the volume is
quite complicated. It is preferable to expand
the first order term, so that, using the identity
(\ref{eq:rela}) and
 (\ref{eq:vol1}),
we have
\begin{eqnarray}
V_{(2)} &=& {\epsilon \over 2} \tilde{V}_{(1)} = {\epsilon \over 2} \left[
\int dA_{(1)} ({\bf W} \cdot
{\bf n})  + \int dA ({\bf W} \cdot {\bf n}_{(1)}) \right]
\nonumber \\
&=& {\epsilon^2 \over 2} \int dA [  ({\bf e}_a \cdot {\bf W}^a )
({\bf W} \cdot {\bf n}) -  ({\bf e}_a \cdot {\bf W} )
({\bf W}^a \cdot {\bf n})]\,.
\label{eq:vol2}
\end{eqnarray}

Let us consider the second order term in the expansion of the total mean curvature $M$.
We have that
\begin{equation}
M_{(2)} = \int [
dA_{(2)} \; K +  dA_{(1)} \; K_{(1)} + dA \; K_{(2)} ]\,,
\end{equation}
so  that using  (\ref{eq:sqrt}) and (\ref{eq:k}), we obtain
\begin{eqnarray} \fl
M_{(2)} &=& \epsilon^2 \int dA \; [
{1 \over 2} ({\bf e}_a \cdot {\bf W}^a )^2 K
- {1 \over 2} ({\bf e}_a \cdot {\bf W}_b) K
({\bf e}^b \cdot {\bf W}^a)
- ({\bf e}_a \cdot {\bf W}^a )
({\bf n} \cdot \nabla_a {\bf W}^a )
\nonumber \\ \fl
&-&
2 K^{bc}({\bf e}_a \cdot {\bf W}^a )
({\bf e}_b \cdot  {\bf W}_c )
+ 2 ({\bf e}^a \cdot {\bf W}^b )
({\bf n} \cdot \nabla_a {\bf W}_b )
+ 2 K^{ab} ({\bf e}_a \cdot {\bf W}_c )
({\bf e}^c \cdot {\bf W}_b )
\nonumber \\ \fl
&+&  K^{ab} ({\bf e}_a \cdot {\bf W}_c ) ({\bf e}_b \cdot {\bf W}^c )
+ ({\bf n} \cdot {\bf W}_c ) ( {\bf e}^c \cdot \nabla_a {\bf W}^a )
- K_{ab} ({\bf n} \cdot {\bf W}^a ) ({\bf n} \cdot {\bf W}^b )]\,.
\label{eq:m21}
\end{eqnarray}
There is no obvious simplification. On the other hand, using the
identity (\ref{eq:rela}) and  (\ref{eq:m11}), we obtain, up to
a total divergence, the simpler expression
\begin{eqnarray}
M_{(2)} = {\epsilon \over 2} \tilde{M}_{(1)}
&=& {\epsilon^2 \over 2} \int dA \; \{ 2 (K^{ab} - K g^{ab})
({\bf n} \cdot \nabla_a {\bf W}_b ) ({\bf n} \cdot {\bf W})
\nonumber \\
&-& {\cal R} [
({\bf n} \cdot {\bf W}_a ) ({\bf e}^a \cdot
{\bf W} ) +
({\bf n} \cdot {\bf W} ) ({\bf e}^a \cdot
{\bf W}_a ) ] \} \,.
\label{eq:m22}
\end{eqnarray}
These two expressions differ only by a total
divergence. However, it is  quite involved to extract it from
(\ref{eq:m21}).

For the bending energy, we have that a direct expansion
is the more convenient approach and with
\begin{equation}
\fl
F_{b\, (2)} = \int [ dA_{(2)} \; K^2  +  2 dA_{(1)} \; K \; K_{(1)}
+ dA K_{(1)} K_{(1)} 
+ 2 dA \; K K_{(2)} ]\,,
\end{equation}
using  (\ref{eq:sqrt}), (\ref{eq:k}), we obtain
\begin{eqnarray} \fl
F_{b\, (2)} &=&
\epsilon^2 \int dA \; [ ({\bf n} \cdot \nabla_a {\bf W}^a )^2
+ 2 (2 K^{ab} - K g^{ab} )  ({\bf e}_a \cdot {\bf W}_b )
 ({\bf n} \cdot \nabla_c {\bf W}^c ) \nonumber \\ \fl
&+& 4 K  ({\bf e}^a \cdot {\bf W}^b ) ({\bf n} \cdot \nabla_a {\bf W}_b )
+ 2 K  ({\bf n} \cdot {\bf W}^c ) ({\bf e}_c \cdot \nabla_a {\bf W}^a )
\nonumber \\ \fl
&-& 2 K K^{ab}  ({\bf n} \cdot {\bf W}_a ) ({\bf n} \cdot {\bf W}_b )
- {1 \over 2} K^2 ({\bf n} \cdot {\bf W}_a ) ({\bf n} \cdot {\bf W}^a )
\nonumber \\ \fl
&+& ( 4 K^{ab} K^{cd} + 4 K K^{ad} g^{bc} - 4 K K^{cd} g^{ab}
+ 2 K K^{ac} g^{bd} + {1 \over 2} K^2 g^{ab} g^{cd}
\nonumber \\ \fl
&-& {1 \over 2} K^2 g^{ad} g^{bc})
({\bf e}_a \cdot {\bf W}_b ) ({\bf e}_c \cdot {\bf W}_d )]\,.
\label{eq:fb22}
\end{eqnarray}

These expressions provide directly the second variation of the
Helfrich-Canham Hamiltonian in terms of the deformation vector
${\bf W}$ and its first and second derivatives.
It is desirable, however, in order
to make contact with the expressions derived in  \cite{Defo}, to decompose
${\bf W}$ into tangential and normal. This involves plugging
 (\ref{eq:c1}) to (\ref{eq:c4}) into the covariant expressions
we have derived.
For example, for  the area functional we obtain
\begin{eqnarray}
\fl
A_{(2)} &=& {\epsilon^2 \over 2} \int dA
[ \nabla_a \Phi \nabla^a \Phi + (K^2 - K_{ab} K^{ab} )\Phi^2
- 2 K^{ab} \nabla_ ( \Phi_b  \Phi ) + K_{ac} K_b{}^c \Phi^a \Phi^b
\nonumber \\ \fl
&+& 2 K \Phi \nabla_a \Phi^a + (\nabla_a \Phi^a )^2
- (\nabla_a \Phi^b ) \nabla_b \Phi^a - 2 K^{ab} \Phi \nabla_a \Phi_b ]\,.
\end{eqnarray}
This is not the most useful form, however. Integrating by parts and isolating a total
divergence,  we can write it down
in an alternative way as
\begin{eqnarray}
\fl
A_{(2)} &=& {\epsilon^2 \over 2} \int dA
\{ - \Phi \nabla^2 \Phi - K_{ab} K^{ab} \Phi^2
\nonumber \\ \fl
&+&
 K ( K \Phi^2 + K _{ab} \Phi^a \Phi^b - 2 \Phi^a \nabla_a \Phi )
\nonumber \\ \fl
&+& \nabla_a [ \Phi \nabla^a \Phi + 2 \Phi  \Phi_b (K g^{ab} - K^{ab} )
- \Phi^b \nabla_b \Phi^a + \Phi^a \nabla_b \Phi^b ]\}\,.
\label{eq:a2c}
\end{eqnarray}
The first line is the normal part of the second deformation. The
second line is proportional to the trace of the extrinsic
curvature $K$. It is essential to recognize that $K$ is the
Euler-Lagrangian derivative for the area functional, $E_A = K$.
Therefore, at equilibrium, the second line will vanish. The third
line is a total divergence and, for a closed vesicle without
boundaries, it can be set to vanish.

This example indicates that it is possible to obtain in a
systematic way simpler expressions by isolating terms that are
total divergences. Let us use the expression (\ref{eq:F1}) for the
first variation of the Helfrich-Canham Hamiltonian (see
\cite{Defo} for an equivalent argument in an alternative
language). Using the identity (\ref{eq:rela}), we have that
\begin{eqnarray}
\fl
F_{(2)} [{\bf X}, {\bf W}] =  {\epsilon \over 2} \; \widetilde{F}_{(1)} [
{\bf X}, {\bf W} ]  &=& {\epsilon \over 2}
\{  \int dA \; E_{F\,(1)} \; ({\bf n} \cdot {\bf W})
+ \int dA \; E_F \; ({\bf n}_{(1)} \cdot {\bf W})
\nonumber \\ \fl
&+& \int dA_{(1)} \; E_F \; ({\bf n} \cdot {\bf W})
+ \int d^2 \xi \; \partial_a [( \sqrt{g} {\cal Q}^a )_{(1)}] \}\,,
\end{eqnarray}
where we have rewritten the second term of  (\ref{eq:F1}) so
that $\sqrt{g} {\cal Q}^a$ is a scalar density of weight one; its
divergence is then independent of the affine connection, so that
variation and derivation commute. $E_{F\,(1)}$ denotes the first
order variation of the Euler-Lagrange derivative appearing in 
(\ref{eq:shape}). We now use  (\ref{eq:sqrt}), (\ref{eq:not})
to obtain
\begin{eqnarray}
\fl
F_{(2)} [{\bf X}, {\bf W}]
&=& {\epsilon^2 \over 2}  \int dA \; E_F \; [
({\bf e}_a \cdot {\bf W}^a ) ({\bf n} \cdot {\bf W})
- ({\bf e}_a \cdot {\bf W} ) ({\bf n} \cdot {\bf W}^a)]
\nonumber \\ \fl
&+& {\epsilon \over 2} \int
dA \; E_{F\, (1)} ({\bf n} \cdot {\bf W})
+ {\epsilon \over 2} \int d^2 \xi \; \partial_a [( \sqrt{g} {\cal Q}^a)_{(1)} ]\,.
\label{eq:f2}
\end{eqnarray}
We now use the results of Sect. 4,  to express in components the
deformations. Furthermore we split the first order correction of
the Euler-Lagrange derivative $E_{F\,(1)}$ in its normal and tangential
parts as $E_{F\,(1)} = E_{F\,(1)
 \, \mbox{\sf perp.}} + E_{F\,(1) \, \mbox{\sf tang.}}
$, where, since $E_F$ is a scalar,  $E_{F\,(1) \, \mbox{\sf tang.}} =
\Phi^a \nabla_a E_F$. It follows that we can rewrite
(\ref{eq:f2}) as
\begin{eqnarray}
\fl
F_{(2)} [{\bf X}, {\bf W}]
&=& {\epsilon^2 \over 2}  \int dA \; E_F \; [ K \Phi^2 + \Phi \nabla_a
\Phi^a - \Phi^a \nabla_a \Phi + K_{ab} \Phi^a \Phi^b ]
+ \Phi \Phi^a \nabla_a E_F \nonumber \\ \fl
&+& {\epsilon \over 2} \int
dA \; E_{F\,(1)\, \mbox{\sf perp}} ({\bf n} \cdot {\bf W})
+ {\epsilon \over 2} \int d^2 \xi \; \partial_a
[( \sqrt{g} {\cal Q}^a)_{(1)} ]\,,
\label{eq:f3}
\end{eqnarray}
or as
\begin{eqnarray}
\fl
F_{(2)} [{\bf X}, {\bf W}]
&=& {\epsilon \over 2}  \int
dA \; E_{F\,(1)\, \mbox{\sf perp}} ({\bf n} \cdot {\bf W})
+ {\epsilon^2 \over 2}  \int dA \; E_F \; [ K \Phi^2  -
2 \Phi^a \nabla_a \Phi + K_{ab} \Phi^a \Phi^b ]
\nonumber \\ \fl
&+& {\epsilon^2 \over 2} \int dA \; \nabla_a ( E_F \Phi \Phi^a )
+ {\epsilon \over 2} \int d^2 \xi \; \partial_a [( \sqrt{g} {\cal Q}^a)_{(1)} ]\,.
\label{eq:f4}
\end{eqnarray}
The second line is a total divergence, and it can be set to vanish
safely. In the first line, the term proportional to the
Euler-Lagrange derivative $E_F$ is surprisingly simple. We
recognize the same structure that appears in the second variation
of the area functional in the form (\ref{eq:a2c}).

When the shape equation $E_F =0$ is satisfied, up to a total
divergence, we have that the second variation is simply
\begin{equation}
F_{(2)} [{\bf X}, {\bf W}]
= {\epsilon \over 2} \int
dA \; E_{F\,(1) \, \mbox{\sf perp.}} ({\bf n} \cdot {\bf W})\,.
\label{eq:f2b}
\end{equation}

If we set the tangential part of the deformation to vanish, $\Phi^a = 0$,
then the the second variation takes the form
\begin{equation}
F_{(2)} [{\bf X}, {\bf W}]
= {\epsilon \over 2} \int
dA \; [ E_{F\,(1) \, \mbox{\sf perp.}}
({\bf n} \cdot {\bf W})
+ \epsilon E_F K ({\bf n} \cdot {\bf W})^2 ]\,.
\label{eq:f2c}
\end{equation}

Let us consider now the expressions in components of the
remaining terms in the second variation of the Helfrich-Canham
Hamiltonian in the form (\ref{eq:f4}). For the volume we have that
\begin{equation}
V_{(2)} = {\epsilon^2 \over 2} \int dA \; [ K ( K \Phi^2 +
K _{ab} \Phi^a \Phi^b - 2 \Phi^a \nabla_a \Phi ) +   \nabla_a (\Phi \Phi^a )]\,.
\end{equation}
This case is quite special, since the contribution of the volume to the
Euler-Lagrange derivative is $E_V=1$, therefore, for the volume, $E_{V\,(1)}
= 0$.

For the total mean extrinsic curvature specializing to components
either (\ref{eq:m21}) or (\ref{eq:m22}), we
obtain, up to a total divergence,
\begin{eqnarray}
M_{(2)} &=& \epsilon^2 \int dA \; [ (K^{ab} - K g^{ab} ) \Phi \nabla_a
\nabla_b \Phi -
{1 \over 2} {\cal R}
K \Phi^2
\nonumber \\
&+& {\epsilon^2 \over 2} \int dA \; {\cal R} ( K \Phi^2 +
K _{ab} \Phi^a \Phi^b - 2 \Phi^a \nabla_a \Phi )\,.
\end{eqnarray}
Note that the second line is proportional to the Euler-Lagrange
derivative for $M$, since $E_M = {\cal R}$.

For the bending energy, a direct specialization to components of
 (\ref{eq:fb22}) produces 74 terms, and it is impossible to
tell the trees from the forest. However, the general
considerations that lead to  (\ref{eq:f4}) imply that the
dependence on the tangential component of the deformation $\Phi^a$
is determined. Therefore, we can keep  only the normal part of
the deformation $\Phi$, so that, up to a total divergence, we
obtain
\begin{eqnarray}
\fl
F_{b \, (2)} &=& \epsilon^2 \int dA \; \{
(\nabla^2 \Phi )^2 + {1 \over 2} ( K^2 - 2 {\cal R})
\Phi \nabla^2 \Phi
+ 2 K K^{ab} \Phi \nabla_a \nabla_b \Phi
\nonumber \\ \fl
&+& K (\nabla_a K ) ( \nabla^a \Phi ) \Phi
- 2 K^{ab} (\nabla_a K ) (\nabla_b \Phi ) \Phi
+ (K^4 - {5 \over 2} K^2 {\cal R} + {\cal R}^2 ) \Phi^2 \} \,.
\end{eqnarray}
This corresponds to the form (\ref{eq:f2c}) of the second
variation.

\section{ Concluding remarks}

We have presented in some detail a fully covariant approach to the
deformations of the Helfrich-Canham Hamiltonian; where
applicable, we have compared it to the approach adopted in \cite{Defo}
in which tangential deformations were discarded.

On balance, we feel that there is something to be learnt from
both approaches; the reader who has considered both may better
judge which form of perturbation theory is more appropriate to
the issue being addressed. For example, whereas it is trivially
obvious in the covariant approach that the metric tensor is
subject to variations of second order and no higher, once the
decomposition has been effected, this fact becomes heavily
disguised and would appear to involve miraculous cancellations.
On the  other hand, the second order variation of the Hamiltonian
about an equilibrium configuration is considerably more
transparent  when expressed in terms of normal deformations.

The nature of tangential deformations of individual geometrical
tensors has been clarified at second order, an issue clearly
beyond the scope of the analysis in \cite{Defo} to address.
We have shown giving  explicit examples that, at this order and
higher, tangential deformations are not reparametrizations. We
suspect that there is much still to be pinned down on the issue.
Even in the well explored field of general relativity,
disentangling coordinate artifacts from physical perturbations at
second order remains a vexed issue \cite{Branden}.

We have not examined explicitly any order in perturbation theory
higher than second. This is not going to be simple; it remains a
highly non-trivial challenge. The systematic approach we have
outlined will, it is hoped, provide a few reliable signposts.

\ack
JG would like to thank Denjoe O' Connor for
hospitality during his stay at DIAS.
Partial support to JG from DGAPA-PAPIIT grant IN114302 is
acknowledged. We acknowledge partial support from CONACyT.
We thank Markus Deserno for useful comments.

\end{document}